\documentclass[fleqn,usenatbib]{mnras}

\usepackage{newtxtext,newtxmath}
\usepackage[T1]{fontenc}

\usepackage[normalem]{ulem}  
\newcommand\redout{\bgroup\markoverwith
{\textcolor{red}{\rule[0.5ex]{2pt}{0.8pt}}}\ULon}

\usepackage{tikz,xcolor,hyperref}

\definecolor{lime}{HTML}{A6CE39}
\DeclareRobustCommand{\orcidicon}{
	\begin{tikzpicture}
	\draw[lime, fill=lime] (0,0) 
	circle [radius=0.16] 
	node[white] {{\fontfamily{qag}\selectfont \tiny ID}};
	\draw[white, fill=white] (-0.0625,0.095) 
	circle [radius=0.007];
	\end{tikzpicture}
	\hspace{-2mm}
}

\foreach \x in {A, ..., Z}{\expandafter\xdef\csname orcid\x\endcsname{\noexpand\href{https://orcid.org/\csname orcidauthor\x\endcsname}
			{\noexpand\orcidicon}}
}

\usepackage{graphicx}	% Including figure files
\usepackage{amsmath}	% Advanced maths commands
\usepackage{pdflscape} % Landscape pages
\usepackage{slantsc}
\title[Planetary nebulae hosting accreting WDs]{Planetary nebulae hosting accreting white dwarfs: A possible solution for the mysterious cut-off of Planetary Nebula Luminosity Function? }

\author[D. Souropanis et al.]{
D. Souropanis,$^{1,2,3\orcidA{}}$\thanks{E-mail: d.souropanis@noa.gr}
A. Chiotellis,$^{1,4\orcidB{}}$
P. Boumis,$^{1\orcidC{}}$
D. Jones,$^{5,6,7\orcidD{}}$
S. Akras$^{1\orcidE{}}$\\
\\
\\
$^{1}$Institute for Astronomy, Astrophysics, Space Applications
and Remote Sensing, National Observatory of Athens,
15236 Penteli, Greece\\
$^{2}$Department of Physics, National and Kapodistrian University of Athens, Panepistimiopolis, 15784 Zografos, Greece\\
$^{3}$ Isaac Newton Group of Telescopes, Apartado 321, E-38700 Santa Cruz de La Palma, Canary Islands, Spain\\
$^{4}$ 4rth Lykeion Acharnon, Echinou and Chalkidos , 136 71   Acharnes, Greece\\
$^{5}$ Instituto de Astrof\'isica de Canarias, E-38205 La Laguna, Tenerife, Spain\\
$^{6}$ Departamento de Astrof\'isica, Universidad de La Laguna, E-38206 La Laguna, Tenerife, Spain\\
$^{7}$ Nordic Optical Telescope, Rambla Jos\'e Ana Fern\'andez P\'erez 7, 38711, Bre\~na Baja, Spain
}

\date{Accepted 2023 February 14. Received 2023 February 14; in original form 2022 October 31}

\pubyear{2022}

\begin{document}
\label{firstpage}
\pagerange{\pageref{firstpage}--\pageref{lastpage}}
\maketitle

\begin{abstract}

Many binary companions to the central stars of planetary nebulae (PNe) are found to be inflated, perhaps indicating that accretion onto the central star might occur during the PN phase. The discovery of a handful of nova eruptions and supersoft X-ray sources inside PNe supports this hypothesis. 
In this paper, we investigate the impact that hosting a steadily-accreting WD would have on the properties and evolution of a PN. 
By pairing the published accreting nuclear-burning WD models with radiation transfer simulations, we extract the time evolution of the emission line spectra and ionization properties of a PN that surrounds a 0.6$\rm M_{\odot}$ steadily nuclear-burning WD as a function of the mass accretion rate.  We find that accreting WDs are able to form very extended, high excitation, [\ion{O}{iii}]-bright PNe, which are characterised by high nebular electron temperatures. Their properties remain almost invariant with time and their visibility time can be much longer compared to PNe powered by single WDs.
We discuss the implications of our findings in explaining specific characteristics observed in PNe. Finally,  we examine how accreting WDs affect the planetary nebula luminosity function (PNLF) by covering WD masses in the range of 0.5-0.8$\rm M_{\odot}$ and for various accretion rates within the steady accretion regime.  We find that for all but the lowest accretion rates, the [\ion{O}{iii}]-luminosities are almost constant and clustered very close to the PNLF cut-off value. Our results suggest that mass-accreting WDs in interacting binaries might play a role in understanding the invariant cut-off of the PNLF.

\end{abstract}


%%%%%%%%%%%%%%%%%%%%%%%%%%%%%%%%%%%%%%%%%%%%%%%%%%

%%%%%%%%%%%%%%%%% BODY OF PAPER %%%%%%%%%%%%%%%%%%

\section{Introduction}\label{sec:intro}

Low- and intermediate-mass stars ($\sim1-8~\rm M_{\odot}$), during the late stages of their evolution, experience extreme mass loss that results in the ejection of their stellar envelopes, leaving only their degenerate cores behind. The exposed core, which will become a white dwarf (WD), is hot enough that it ionizes the surrounding ejecta, which  becomes visible as a planetary nebula (PN) for thousands of years, before dispersing into the interstellar medium \citep[][]{1983Kwok}. While PNe offer unique insights into stellar evolution, gas dynamics, and chemical enrichment of the interstellar medium, their nature and formation mechanism have not been adequately explained.

The spectacular array of PN morphologies,  chemical abundances and ionization properties cannot be fully understood in a single star scenario. Instead, stellar duplicity seems to play an important role in the formation and evolution of PNe, given the high binary fraction of PNe central stars \citep[e.g.][]{2009Miszalski, 2015DeMarco, Jacoby2021} and the theoretical association of  binary evolution paths as a route towards resolving several issues that concern the nature of these celestial nebulae,  such as the complex non-spherical shapes that the majority of PNe display  \citep[][]{2014Garcia,2017Jones}, the abundance discrepancy problem  \citep{Corradi2015,Jones2016,Wesson2018} and the inconsistency between the predicted and observed number of Galactic PNe \citep[][]{2006DeMarco}.

Binary progenitor systems have also been suggested as a possible solution for the long standing problem that regards the $[\ion{O}{iii}]~ \lambda5007$ (hereafter [\ion{O}{iii}]) planetary nebula luminosity function \citep[PNLF;][]{2016Ciardullo, Davis2018}. The [\ion{O}{iii}] PNLF describes the distribution of PNe in a galaxy as a function of their absolute magnitude in the [\ion{O}{iii}] nebular emission line, which depends on the age of the parent stellar population and to less important degree on the metallicity \citep[e.g.][]{1980Jacoby, Dopita1992}. While it changes between different type of galaxies, the PNLF shows a well-defined cut-off value of $M_{5007}=-4.54 \pm 0.04$ \citep{2013Ciardullo}, which is  almost invariant across a very broad range of galaxy types and stellar populations. For that reason, the PNLF has been established as an important extragalactic distance estimator, but there is no solid theory that explains why it should work in such a wide array of galaxies and stellar populations. Recently, \citet{2018Gesicki}  claimed that the PNLF invariance among different types of galaxies can be achieved thanks to the new post-AGB
evolutionary models of  \citet{2016bertolammi} that allow lower mass PN central stars to achieve higher luminosities than previously believed.  However, that work has still some important issues especially for the very oldest populations, as in their models the bright cut-off is always the result of progenitors with initial mass in the range 1.1–2 $\rm M_{\odot}$, which are probably a bit massive for that kind of populations. In addition,  \citet{Davis2018}, studied the extinction that affects PNe in M31’s bulge  and showed that even with the most recent evolutionary models of \citet{2016bertolammi}, the most luminous PNe require  central star masses in excess of 0.66 $\rm M_{\odot}$ and  main-sequence progenitors of at least 2.5 $\rm M_{\odot}$, something that is inconsistent with the stars of  M31’s bulge.

Many other alternative scenarios have been proposed over the years attempting to explain the  PNLF cut-off invariance, putting binary evolution as a likely solution to this paradox.  Blue stragglers, common envelope episodes or symbiotic stars are few of the numerous possible  binary systems/processes proposed in the literature  \citep[e.g.][]{2005Ciardullo, Soker2006, Davis2018}. However, it is not clear whether these alternatives are able to explain the puzzling constant cut-off [\ion{O}{III}] 5007 \AA~ luminosity.

%%%%%%%%%%%%%%%%%%%%%%%%%%%%%%%%%%%%%%%%%%%%%%%%%%%%%%%%%%%%%%%

\begin{figure}
\includegraphics[trim=0 0cm 0 0, clip=true,width=\columnwidth,angle=0]{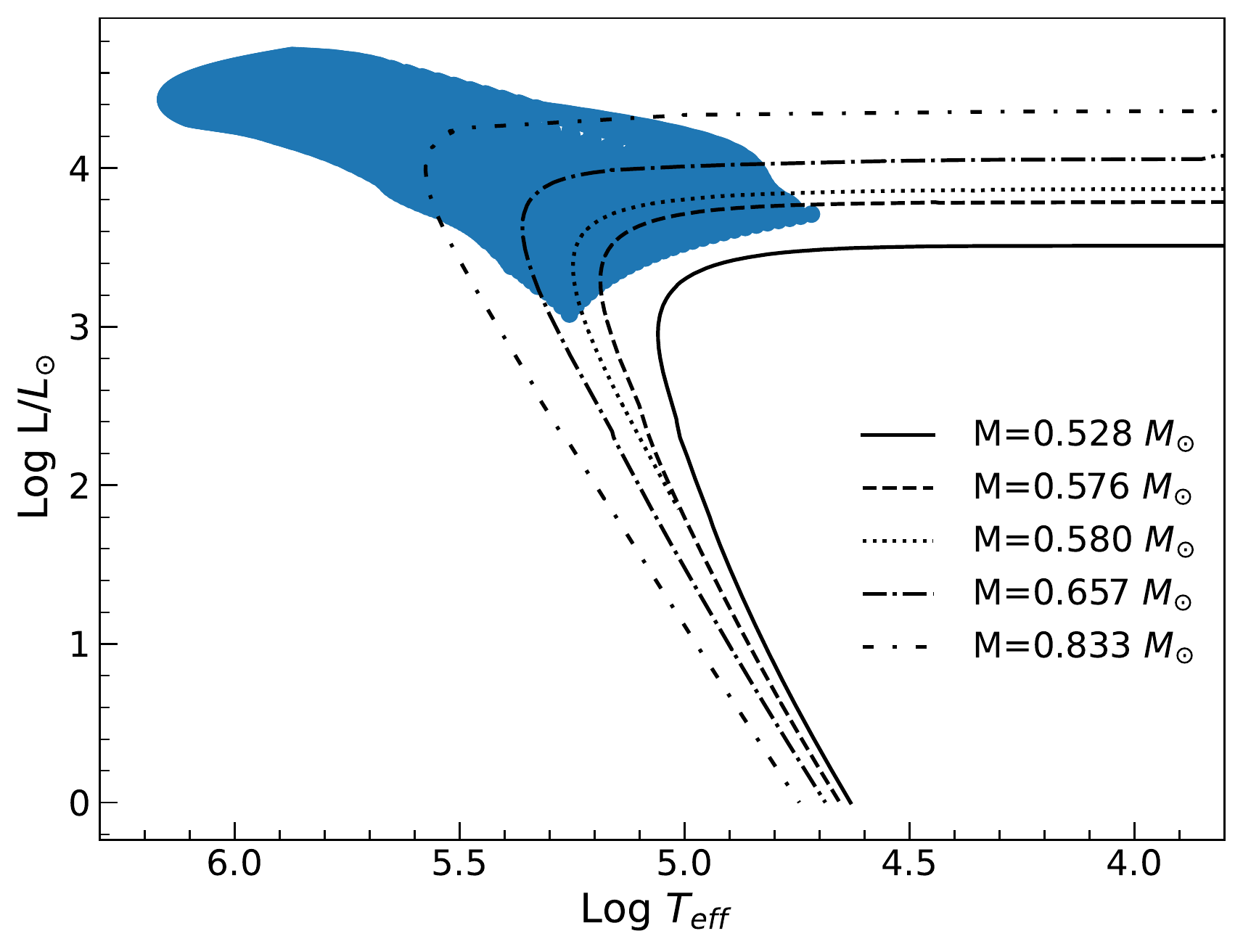} 
\caption {The Hertzsprung-Russell diagram of the post-AGB evolutionary tracks of \citet{2016bertolammi} for different masses and Z=0.02 metallicity. The blue shaded area corresponds to the loci occupied by  hydrogen steadily nuclear burning accreting WDs as extracted by the models of \citet{Nomoto2007}.}
\label{fig:HR_WDs}
\end{figure}

%%%%%%%%%%%%%%%%%%%%%%%%%%%%%%%%%%%%%%%%%%%%%%%%%%%%%%%%%%%%%%%

Taking into account that binary stars seem to be the majority of central stars of PNe (CSPNe) and are placed as the key solution in several problems that concern the nature of PNe, reasonably it raises  the question whether -and to which extent- these binaries are interacting and if this is the case, which would be the characteristics of the hosting PNe.

This question is not rhetorical, as several direct and indirect observational evidence indicate that mass accretion takes place in several binary central stars of PNe. For instance, binary systems related to accreting WDs such as classical novae \citep{1987Bode, 2008Wesson}, symbiotic stars  \citep{2004Guerrero,2004Santander, 2011Corradi, 2013Munari,2018Kiewicz,Akras2019} and supersoft X-ray sources \citep{2008Kahabka, 2001Hutchings,2010Mereghetti,2022Maitra}  have been observed to be hosted in the centres of PNe.  Mass accretion processes have also been suggested to take place at the central stars of the PNe N66 \citep[][]{2003Hamann} and the Eskimo Nebula \citep[][]{2019Guerrero} as a plausible explanation for the two optical outbursts and the variable X-ray emission observed at the centre of the two nebulae, respectively.  The discovery of short-period binaries such as PN M3-1, whose components are close to Roche lobe filling – comprise an additional vigorous argument that central binary interactions can occur before the surrounding PN will be fully dissipated \citep{2019Jones}. Finally, indirect evidence comes from observations of Type Ia Supernovae (SNe Ia) -i.e. the supernovae that result from the thermonuclear combustion of carbon-oxygen WDs in interacting binary systems.  An increasing number of papers argue that many SNe Ia occur in and subsequently interact with PNe, based on polarized light that  SNe Ia reveal  \citep{Cikota2017} and the morphological and dynamical properties of their supernova remnants \citep[e.g.][]{Tsebrenko2013,Tsebrenko2015, 2020Chiotellis,2021Chiotellis}.

An intriguing fact is that mass accreting WDs possess completely different properties and evolution than the single ones. The reason for this is the different energy deposits that the two classes of WDs use for their electromagnetic radiation. Contrary to isolated WDs that radiate their internal energy and slowly cool down, the radiation of accreting WDs is stimulated by the surface nuclear burning that in turn is triggered by the mass accretion. Consequently, their radiation properties are dominantly determined by the WD mass and the accretion rate, and steadily accreting WDs occupy a much broader region on the Hertzsprung-Russell diagram being overall more luminous and hotter sources as compared to their isolates counterparts (see Figure~\ref{fig:HR_WDs}).  In addition, given that accreting WDs can sustain their radiation as long as mass accretion occurs, they can be more persistent radiation/ionization sources than single WDs, having lifetimes that can last up to several Myrs. 

The substantial differences that the two classes of WDs display, are expected to be reflected on the properties and evolution of their surrounding PNe.  Particularly, PNe that host steadily accreting WDs, enclose more efficient and tenacious ionization sources, and hence, they are expected to be characterized by higher ionization/excitation states while their visibility time window can be much broader.   

Despite the existence of the above observational data and theoretical arguments, to date there has been no thorough modelling of PNe that host steadily accreting WDs at their centres. Through this work we aim to conduct the first one.  Specifically, by coupling the known WD accretion models of \citet{Nomoto2007} with radiation transfer and photo-ionization numerical techniques, we study the properties and evolution of  a PN that surrounds a 0.6~$\rm M_{\odot}$ steadily accreting WD, setting as a free parameter the mass accretion rate. For comparison, we run an additional set of simulations but this time considering the evolutionary sequence of a single WD with
metallicity Z = 0.02 and final mass of 0.58~$\rm M_{\odot}$  WD from the most recent accelerated-evolution models  of \citet{2016bertolammi} and we extract the distinctive differences between the PNe formed by the two classes of objects. Finally, we discuss the implications of our findings in explaining  the observables we receive from PNe, emphasizing on the long-standing and puzzling problem of the [\ion{O}{III}] PNLF constant cut-off. 

The paper is organized as follows:  In Section \ref{sec:2},  we  describe the main principles and assumptions of our modelling as well as the initial set up of the simulations.  The results of our modelling and their analysis are presented in Section \ref{sec:3}. Finally, we discuss our results, and we extract main conclusions in Section \ref{sec:4}.

\section{Methodology}\label{sec:2}

We model the time evolution of the emission line spectra of PNe that host steadily accreting WDs at their centres using the accreting WDs models of \citet{Nomoto2007} and the photoionization code {\sc cloudy} \citep[v17.02,][]{Ferland2017}. {\sc cloudy} is an non-local thermodynamic
equilibrium (NLTE) spectral synthesis and plasma
simulation code designed to simulate conditions in interstellar matter under a broad range of conditions (i.e. gas density, composition, incident spectrum etc). More details, source files and all necessary data are available from \url{www.nublado.org}. We choose a WD with mass of 0.6 $ \rm M_{\odot}$ that accretes H-rich material at its minimum  ($\dot{M}_{\rm st}= 3.3 \times 10^{-8}~\rm M_{\odot}~yr^{-1}$) and its maximum ($\dot{M}_{\rm cr}= 1.2 \times 10^{-7}~\rm M_{\odot}~yr^{-1}$) accretion rate within the steady accretion regime (we refer to \citet{2022Souropanis} and \citet{Nomoto2007} for more details of the models).  We run a set of photoionization models for each context by evolving the gaseous shell but keeping fixed the luminosity and temperature of the ionizing source. This is because the temporal behaviour of the photoionisation rates of  WDs that accrete mass with constant accretion rates  is not expected to change substantially within the short-lived PN phase.  

In order to clarify how the observables of PNe that host accreting WDs differ from those powered by single WDs,  we run in parallel a set of photoionization models for the post-AGB evolutionary track for a WD with mass of 0.58 $\rm M_{\odot}$ and metallicity of Z=0.02 as computed by \citet{2016bertolammi}. The model corresponds to the evolution of a progenitor with initial mass of 2 $\rm M_{\odot}$.
This time we allow the modelled gaseous shell to evolve together with the central
star, for which the temperature and luminosity change with age according to the evolutionary model.  The ionizing continuum for all cases is assumed to be a blackbody as it has the virtue of simplicity and provides a reasonable approximation of the ionizing emission of nuclear-burning accreting WDs and single WDs, except far into the Wien tail \citep{Chen2015}.

To simulate the evolution and the structure of the nebular shell, in our models we  assume spherical symmetry. The inner shell radius is set to expand at a constant velocity of 25 $\rm km~ s^{-1}$ while the nebular density (constant across the nebular shell) decreases with the square of time. The initial inner radius and nebular number density are set to $\rm 10^{16.8}~cm$  and  $\rm 10^{5.5}~cm^{-3}$, respectively,  considering as the starting point of our simulations the moment where the central source has a temperature of 30~000~K as extracted by the evolutionary track of 0.58~$\rm M_{\odot}$, Z=0.02  model \citep[][]{2016bertolammi}. We terminate our calculations when the gas temperature
drops below 4000 K. The nebular abundances were taken from \citet{1983Aller} and \citet{1989SKhromov} and no dust presence was assumed. These parameters were chosen to result in ionization bounded (optically thick) nebulae and were adopted for each of the considered by us central star models. Finally, we constrain our modeling to 7000 years, as at later stages of the evolution of the shell, which becomes very dispersed, the total nebular ionized masses in order to result in ionization bounded PNe increase a lot taking values that are not within reasonable/conservative limits (0.01 and 3  $\rm M_{\odot}$) and the ionization bounded approximation is not anymore valid \citep{2008Frew}.

\section{Results}\label{sec:3}

\subsection{Optical properties of the planetary nebulae hosting accreting white dwarfs}

\begin{figure*}
\includegraphics[trim=0cm 0cm 0cm 0cm, clip=true,width=\textwidth,angle=0]{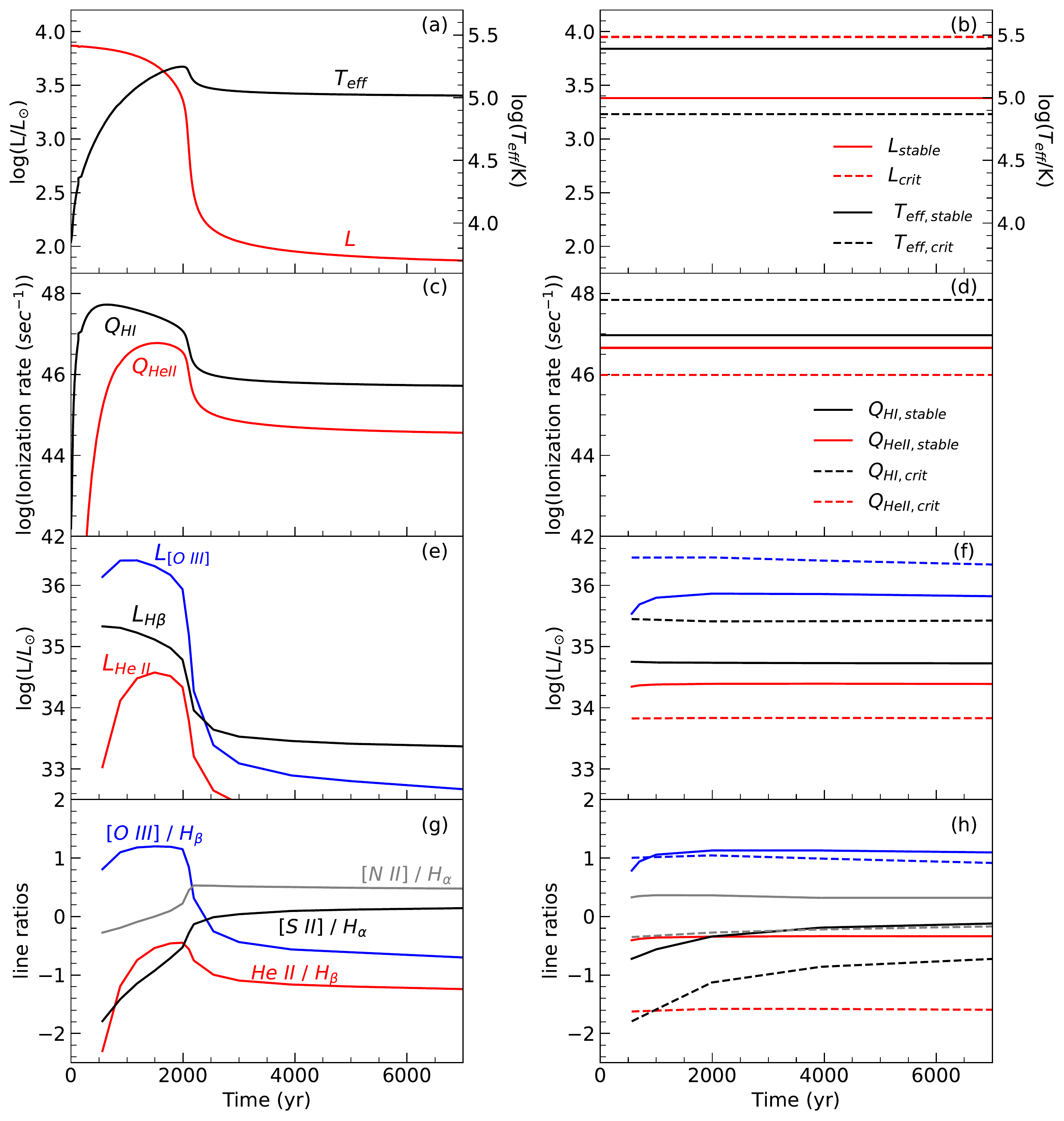} 
\caption {The time evolution of the central star's luminosity and effective temperature (a,b),  H~\textsc{I} and He~\textsc{II} ionizing photon rates (c,d), nebular total line luminosities (e,f) and line rations (g,h) of a single 0.58~$\rm M_{\odot}$ WD (left column) and a steadily accreting 0.6~$\rm M_{\odot}$ WD (right column) that accretes H-rich matter at its minimum and maximum accretion rate within the steady accetion regime. Note also that in all panels in the right column the different line style refers to the mass accretion rate, where the maximum is presented with a dashed line and the  minimum with a solid line, respectively. The line ratios are in log scale. See more details in the text. }
\label{fig:neb_lines}
\end{figure*}

Employing the methodology described in Section \ref{sec:2}, we extract the time evolution of the line emission spectrum of PNe powered by a 0.6~$\rm M_{\odot}$ accreting WD and how it differs from those powered by a single one  of roughly equal mass. The sequence of panels in Figure~\ref{fig:neb_lines} illustrate the time evolution  of the luminosity, effective temperature (plots a,b) and the corresponding photon rates capable to ionize  H~\textsc{I} and He~\textsc{II} (c,d) for the assumed central star, as well as, the time evolution of the nebular line luminosities (e,f) and line ratios (g,h) at the wavelengths of a few important lines typically observed in PNe.
 
As shown in Figures \ref{fig:neb_lines}a,c, the temporal behaviour of the ionisation rates for a single WD is determined by the variation of its effective temperature ($\rm T_{eff}$) and luminosity (L) that follow the evolution of the central star. During the early evolutionary phase, the ionisation rates significantly increase as the star heats up at nearly constant luminosity, reach a peak and then rapidly decrease -by one to two orders of magnitude- as the WD enters the cooling track.

This is not the case for the steadily accreting WDs (Figure \ref{fig:neb_lines}b,d) as their intrinsic luminosity and effective temperature are dominantly determined  by the mass accretion and thus, they remain very strong  sources of ionizing radiation at all stages of planetary nebula evolution. As illustrated in Figure \ref{fig:neb_lines}b the WD's luminosity rises by about a factor of four  with increasing the accretion rate from its minimum ($\dot{M}_{\rm stable}$) to the maximum value ($\dot{M}_{\rm crit}$) as the accreted material is burned at higher rates. By contrast, the effective temperature  decreases, due to the expansion of the nuclear burning WD’s photospheric radius. These temperature and luminosity variations directly affect the intensity and hardness of the nuclear burning WD’s radiation field, whereas it is evident from Fig. \ref{fig:neb_lines}d the number of H~\textsc{I} ionizing photons rises with increasing the accretion rate. This can be understood as the result of two effects that counteract each other. As we have discussed previously, WDs accreting at higher rates, although their effective temperatures are decreased, have higher luminosities, and hence, more inflated photospheres. The higher luminosity favours the production of ionizing photons, while the lower temperature disfavours it. The net effect is a mild increase of ionizing photons with increasing the accretion rate. This trend breaks for  the flux of He~\textsc{II} ionizing photons, which decreases with increasing the accretion rate. This is because the colour temperatures of low-mass rapidly-accreting WDs can fall below $10^{5} \rm K$, and thus, they are incapable of producing an appreciable number of He~\textsc{II} ionizing photons. Overall, the flux of H~\textsc{I} and He~\textsc{II} ionizing photons of accreting WDs are strongly enhanced and are about 1-2 orders of magnitude above the estimates for the H~\textsc{I} and He~\textsc{II} ionizing photon fluxes of a 0.58~$\rm M_{ \odot}$ single WD after entering the WD cooling track.

Regarding the different features that may be present in PNe hosting single and accreting WDs at their centres, we studied the evolution of the total nebular luminosities at the wavelengths of a few important lines typically observed in PNe; namely the recombination lines $\rm H{\rm \beta}$ 4861 \AA~ and \ion{He}{II} 4686 \AA~, and the nebular forbidden line [\ion{O}{III}] 5007 \AA (see Figure \ref{fig:neb_lines}e,f).  From our simulations we obtain the volume emissivity of each line, which is then integrated over the simulated volume to get the total line luminosity, as also elucidated in \citet{2022Souropanis}.

Figure \ref{fig:neb_lines}e depicts the evolution of the nebular line luminosities of a single 0.58 $\rm M_{\odot}$ WD for a total period of 7000~yrs. As can be easily noticed, after the early evolutionary phase (at age of $\sim 3000$~yrs) a decrease of all lines luminosities is clearly observed and  reflects the drop of the luminosity of the central star. 
As regards the evolution of the nebular line luminosities in PNe powered by accreting WDs (Fig. \ref{fig:neb_lines}f), we do not detect  significant variations with time.  The [\ion{O}{III}] luminosity possess a very moderate decrease with time, while those of the \ion{He}{II} and $\rm H{\rm \beta}$ remain almost constant. This result indicates that the final outcome of PNe that host accreting WDs is dominantly determined by the accretion properties of their central binary and to a very lesser degree to the position and density of the emitting shell. Thus, very evolved and dispersed PNe that host accreting WDs remain optically bright, possessing line luminosities that can be about 1 to 4 orders of magnitudes higher than their counterparts which surround single WDs. During their whole evolution, the most prominent and brightest line is the [\ion{O}{III}] line. Finally, the   nebular line luminosities are strongly affected by the  mass accretion rate, giving rise of almost 1 order of magnitude when the accretion rate increases from its minimum to its maximum possible value. Exception to this trend is the line of \ion{He}{II} 4686 \AA~ for which the increase of the mass accretion rate is accompanied by a modest decrease of its luminosity.

Except for the nebular line luminosities, a number of line-intensity ratios is also a noticeably different in a PN surrounding an accreting WD as compared to those powered by a single one.  Figures \ref{fig:neb_lines}g,h  illustrate the striking differences of the evolution of the [\ion{O}{III}]~5007 \AA/$\rm H{\beta}$, [\ion{N}{II}]~6584 \AA/$\rm H{\alpha}$, \ion{He}{II}~4686~ \AA/$\rm H{\beta}$ and [\ion{S}{II}]~6717,31 \AA/$\rm H{\alpha}$ line flux ratios in PNe between the single and the accreting WD cases.

The PNe hosting accreting WDs at their centres have [\ion{O}{III}]~5007 \AA/$\rm H{\beta}$ line ratios that are, in general, much greater than the ones resulting from the evolution of a single WD. Particularly, at late times the [\ion{O}{III}]~5007~\AA/$\rm H{\beta}$ ratio can be 4 to 6 times more enhanced. The nebular flux ratios around single WDs in the lower ionization lines [\ion{N}{II}] and [\ion{S}{II}] relative to $\rm H{\rm \alpha}$, are a bit larger than those obtained by accreting WDs at all epochs except the early evolutionary phases where the single star heats-up towards the turn-around point in the Hertzsprung-Russel diagram. The variability of the \ion{He}{II}~4686 \AA/$\rm H{\beta}$ emission line ratio from accreting WDs cannot be firmly drawn, as it strongly depends on the WD's accretion rate, but overall it reveals a much wider range of values that can be significantly smaller or larger than those powered by a single WD.

%\textbf{ Examples include \ion{He}{II}~4686~\AA/$\rm H{\beta}$ $\langle$0.46, 0.026 vs. 0.068$\rangle$, [\ion{N}{II}]~6584 \AA/$\rm H{\alpha}$ $\langle$ 2.08, 0.60 vs. 3.19$\rangle$, [\ion{O}{III}]~5007~\AA/$\rm H{\beta}$ $\langle$13.46, 9.79 vs. 0.27$\rangle$ and [\ion{S}{II}]~6717,31~\AA/$\rm H{\alpha}$ $\langle$0.65, 0.14 vs. 1.24$\rangle$ where the number in brackets are for the accreting WD model that is mass-accreting with its minimum and its maximum accretion rate and the single WD model at an age of 3935 years, respectively.  }

\begin{figure*}
\includegraphics[trim=0cm 0cm 0cm 0cm, clip=true,width=\textwidth,angle=0]{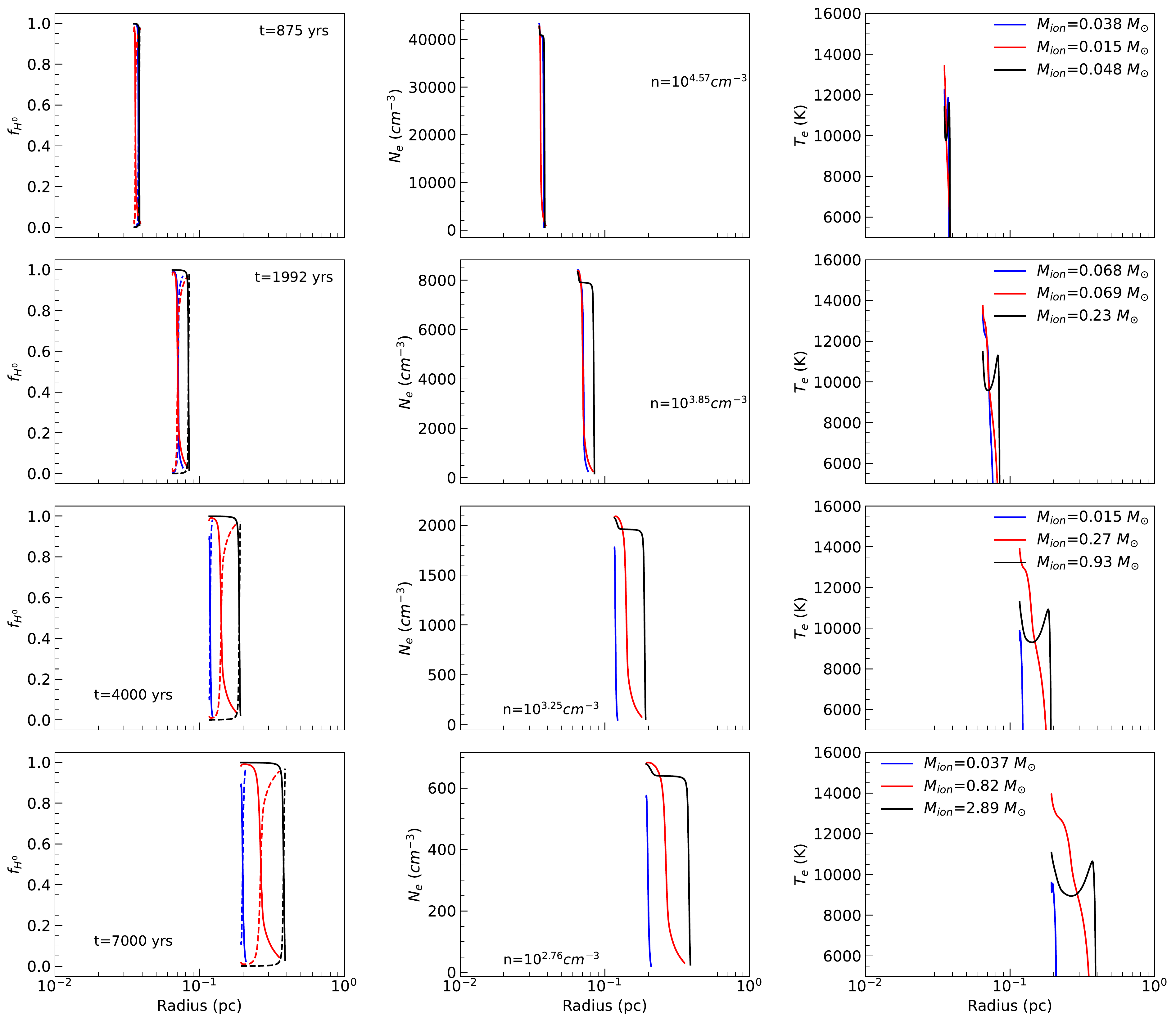} 
\caption {The evolution of the H ionization profiles of the shell (left) and the radial variation of the nebular electron densities (middle) and temperatures (right) as coupled to a 0.58 $\rm M_{\odot}$ single WD model (blue solid line) and to a 0.6 $\rm M_{\odot}$ WD that accretes matter at its minimum (red solid line) and its maximum (black solid line) accretion rate within the steady accretion regime.}
\label{fig:PN_evol}
\end{figure*}

\begin{figure*}
\includegraphics[trim=0 0cm 0 0, clip=true,width=\textwidth,angle=0]{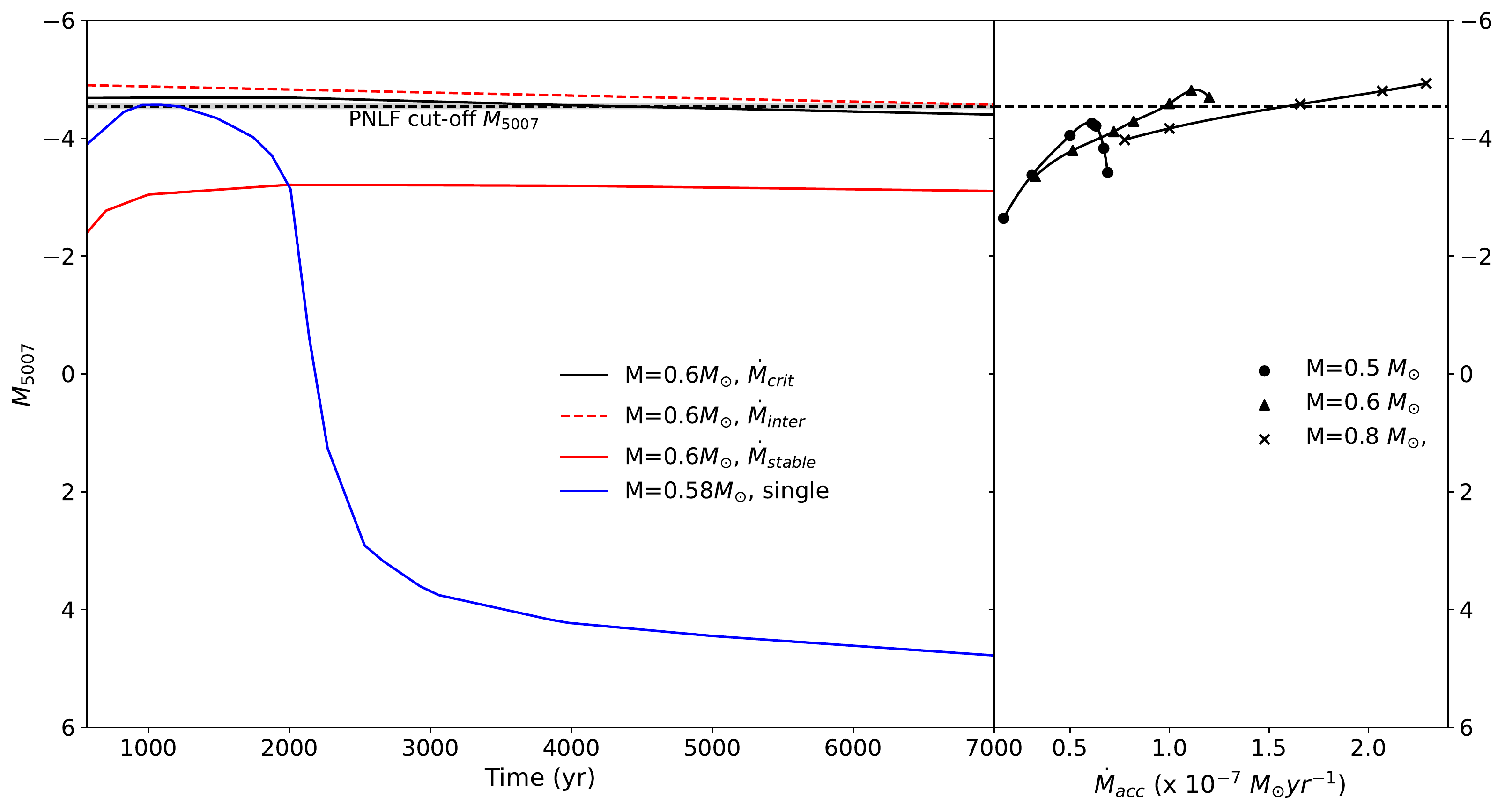} 
\caption {Left: The time evolution of the  [\ion{O}{III}] 5007~\AA~ absolute magnitude for a PN hosting a 0.58~$\rm M_{\odot}$ (blue solid line) and a 0.6~ $\rm M_{\odot}$ accreting WD that accrete H-rich  material with an accretion rate of $\dot{M}_{\rm stable}= 3.3 \times 10^{-8}~\rm M_{\odot}~yr^{-1}$ (red solid line), $\dot{M}_{\rm inter}= 1.1 \times 10^{-7}~\rm M_{\odot}~yr^{-1}$ (red dashed line) and $\dot{M}_{\rm crit}= 1.2 \times 10^{-7}~\rm M_{\odot}~yr^{-1}$ (black solid line).  The black-dashed horizontal line corresponds to the latest estimation of the  [\ion{O}{III}] PNLF cut-off value: $M_{5007}=-4.54 \pm 0.04$ \citep{2013Ciardullo}. Right: The  [\ion{O}{III}] 5007~\AA~ absolute magnitude of PNe that host accreting WDs as a function of the WD mass and accretion rate.  }
\label{fig:oiii}
\end{figure*}

\subsection{Ionization structure, electron temperatures and densities}\label{sec:3.2}

In order to provide a framework useful for interpreting
the general characteristics of planetary nebulae hosting accreting WDs at their centres, we illustrate in Figure \ref{fig:PN_evol} a series of evolutionary
snapshots for PNe models,  consisting of the same set of parameters regarding the evolution of the envelope (see Sect. \ref{sec:2}) but different combinations of central-star models. 

In particular,  Fig. \ref{fig:PN_evol}  depicts the evolution of the H ionization profiles of the shell (left) and the radial variation of the nebular electron densities (middle) and temperatures (right) as coupled to a 0.58~$\rm M_{\odot}$ central star model (blue solid line) and to a 0.6 $\rm M_{\odot}$ WD that accretes matter at its minimum (red solid line) and its maximum (black solid line) accretion rate within the steady accretion regime. In each  evolutionary snapshot, the time refers to the time elapsed from the zero point defined at log$\rm T_{\rm eff}$ = 3.85, where the fast part of the post-AGB evolution towards the white-dwarf domain has started considering the evolutionary sequence with the M=0.58~$\rm M_{\odot}$  model \citep[][]{2016bertolammi}. Fig. \ref{fig:PN_evol} also displays for each of the three WD model and for each evolutionary snapshot our calculations of the total PN ionized mass.

As can be seen in Fig.\ref{fig:PN_evol}, during the early evolutionary stages (t~=~ 875  years), the  differences in the H ionization profiles and the radial variations and structures of electron temperature and density of the shell accounted for both single and accreting WDs are not substantial. This is explained by the similar central stars' ionizing radiation fields at that stage of  their evolution. However, after t=1992 yr, which illustrates the situation when the single central star has reached its maximum surface temperature before it fades away towards the WD's cooling tracks, the differences become essential. Particularly, we find that the radial extent of the ionized medium in our accreting WD models is about 1.4 to 2 times larger than that of the single  WD model at almost all epochs of the evolution. The difference becomes even more discernible during the late evolutionary stages (t= 7000 years), where the sizes of PNe powered by accreting WDs can be up to 0.5 parsec, notably greater than that of the single 0.58~$\rm M_{\odot}$ WD model, as the central star has faded.  Finally, concerning the radial extent of the ionized nebulae around accreting WDs, it is predicted an increase of the size by a factor of 1.3-1.5 when the accretion rate increases from its minimum value to the maximum one.

 The higher ionization efficiency of the accreting WDs as compared to the single one, directly affects the electron density distribution within the ionized shell (middle column of Fig. \ref{fig:PN_evol}) that follows a similar pattern with those of ionized hydrogen radial distribution. Overall, the shell around  accreting WDs  is characterized by higher electrons densities, distributed in a wider ionized shell and extended to larger radii from the central star. These quantities are maximized for the case of the high accreting WD model (black line).

Regarding the electron temperature profiles of PNe powered by accreting WDs, (right column of Fig. \ref{fig:PN_evol}) we find that qualitatively, they follow the same trends at all evolutionary snapshots, except for a radial scale factor. For WDs accreting with low accretion rates within the steady accretion regime, the temperature drops almost monotonically to $10^{4}$ K within a distance which is determined mainly by the assumed density and inner radius of the shell at each evolutionary snapshot. For rapidly accreting WDs, the electron temperatures are lower and actually fall down, reach a minimum and then increase with radial distance out to the ionization front. The same pattern is also observed in the electron temperature profiles of model PNe powered by the 0.58 $\rm M_{\odot}$ single WD. The behaviour of the electron temperature within the ionized region reflects the evolution of the ionizing properties of the central star. In particular, during the early evolutionary stages as the star becomes hotter and its photons more energetic,  the shell's electron temperature increases rapidly and reaches values as high as 15000 K  when the central star is hot enough to ionize helium twice. Subsequently, when the luminosity and effective temperature decrease once the central-star begins to fade rapidly after entering the WD cooling track (at an age of about 2544 years), the electron temperature in the nebula drops accordingly. This is clearly seen in Fig. \ref{fig:PN_evol} at the evolutionary snapshots greater than 1992 years, where the mean temperature of the gas cannot exceed 10\,000~K. In contrary, PNe powered by accreting WDs can maintain high nebular electron temperatures for several thousand of years that can reach up to 12\,000 K and 15\,000 for high and low accretion rates, respectively. PNe powered by accreting WDs demonstrate a positive correlation between the ionised
mass and the nebular size, which is shown in Fig. \ref{fig:PN_evol} . At late stages, the ionized mass increase sharply as the nebula expands and its density decreases. This is not the case for the PNe powered by the single WD model, as due to the decrease of the ionising flux from the
central star, the ionised mass starts to decrease as well,
and correspondingly the ionisation front recedes inside the shell. However, at later ages as the PN density reach lower values, the ionized mass increases again.

\subsection{\texorpdfstring{$[\ion{O}{iii}]$ 5007~\AA}~  absolute magnitudes for PNe hosting accreting WDs}

Given that steadily accreting WDs sustain continuous nuclear burning on their surfaces, have luminosities and effective temperatures appropriate for the formation of [\ion{O}{III}]-luminous PNe and are expected to be hosted in galaxies of both young and old stellar populations, it is worth  investigating their  [\ion{O}{III}] 5007 \AA~ emission of their hosted nebulae and to compare them to the observables we receive regarding the [\ion{O}{III}] PNLF. To do so, we calculate the flux evolution in the [\ion{O}{III}] emission line from each one of the accreting WDs models as described in Section \ref{sec:2} and  subsequently we convert them to absolute magnitudes through the relation: $M_{5007}=-2.5log(F([\ion{O}{III}]))-13.74$; where F([\ion{O}{III}]) is the line flux in units of $\rm erg s^{-1} cm^{-2}$, assuming a distance of 10 pc. Similarly, we  derive the [\ion{O}{III}] magnitudes from the  single evolutionary track with mass of 0.58 $\rm M_{\odot}$.

The PN [\ion{O}{III}] magnitude of the single WD (see Figure \ref{fig:oiii}a) vary significantly as the central star evolves, firstly increasing, reaching the PNLF cut-off value and then quickly starts decreasing as the star enters the cooling track. The time spent at the maximum brightness is insignificant in comparison with the average life-time of PNe.  In the same figure, we also plot the [\ion{O}{III}] nebular magnitude  for the 0.6 $\rm M_{\odot}$ accreting WD model that accretes matter at the two corresponding limits of accretion rate within the steady accretion regime. Finally, a third case has been included in our study for which the 0.6 $\rm M_{\odot}$ accreting WD produces the brightest [\ion{O}{III}] PN around it. This case corresponds to an accretion rate of $\dot{M}_{\rm accr}= 1.1 \times 10^{-7}~\rm M_{\odot}~yr^{-1}$.

Rapidly accreting WD model shows an [\ion{O}{III}] brightness very close to the observed cut-off value at all epochs. Same applies for the WD that accretes with $\dot{M}_{\rm accr}= 1.1 \times 10^{-7}~\rm M_{\odot}~yr^{-1}$ for which the [\ion{O}{III}] brightness remains equal or even higher than the cut-off value through the whole PN evolution. This result indicates that not only young but also very old optically thick planetary nebulae can reach that value, depending on when the accretion phase started.  On the other hand, WDs accreting with low accretion rates reach magnitudes which are around 1 mag below the observed PNLF cut-off value. This is because the ionisation rate for $\rm O^{++}$, that expresses the number of available ionising photons emitted by the central star per unit time falls down with decreasing the accretion rate. In our modeling the resulting hydrogen number densities decrease with post-AGB time from $\sim$ $ 10^{5.5}$ to $\sim$ 500~ $\rm cm^{-3}$. However, for rapidly accreting WDs and low density PNe ($\sim$ 7000 years), due to the high ionizing flux from the source  the maximum PN ionized mass  in order to result  in ionization bounded nebulae reach values as high as 2.89 $\rm M_{\odot}$, which is quite a high but still feasible mass for a PN (see section \ref{sec:3.2}). To better understand the possible observational consequences of reduced PN masses, we have rerun our standard model but keeping constant the total nebular mass for four different cases. Qualitatively, we found that the absolute [\ion{O}{III}] magnitude is reduced by a factor of 1.08, 1.14, 1.25 and 1.39, when the assumed total PN mass is  2.5, 2.0, 1.5 and 1 $\rm M_{\odot}$, respectively.

To further investigate if the PNLF cut-off value could be attained by different WD masses and accretion rates, we calculate the nebular [\ion{O}{III}]  absolute magnitude powered by WD masses ranging from 0.5 to 0.8~$\rm M_{\odot}$ that accrete hydrogen-rich matter at  various accretion rates ($ \Dot{M}_{\rm accr}$) as have been extracted by the models of \citet{Nomoto2007}. For our simulations we assumed a  non-evolving optically thick, constant density shell, with a fixed inner radius of $\rm 10^{17} cm$ and a density of $10^{4.1} \rm cm^{-3}$, which are within the typical range of values for unevolved PNe. Note that as we have shown in Figure \ref{fig:oiii}a the [\ion{O}{III}] magnitudes of our accreting WD models with an evolving gaseous shell do not change significantly as the shell evolves. 
 
For this set of models, Figure \ref{fig:oiii}b presents the resulting [\ion{O}{III}] absolute magnitudes as a function of WD mass and accretion rate within the steady accretion regime. As can be easily noticed,  all steadily accreting WDs can produce very bright [\ion{O}{III}] PNe with magnitudes very close to the observed PNLF cut-off value, exceeding or falling behind it by only a fraction of a magnitude depending on the WD mass and the accretion rate. For the most massive WDs in our models ($M_{\rm WD}= 0.8 \rm M_{\odot}$), the [\ion{O}{III}] brightness rises  monotonically with increasing the accretion rate and for high accretion rates ($\Dot{M}_{\rm accr}$ $\geqslant 1.65 \times 10^{-7}~\rm M_{\odot}~yr^{-1}$) reaches magnitudes above the PNLF cut-off value that can be as high as -4.9 mag. Regarding the lower WD masses ($M_{\rm WD}= 0.5 - 0.6 \rm M_{\odot}$), we  notice that this trend breaks for rapidly accreting WDs  with accretion rates very close to the maximum stable burning rate and the most bright PNe are powered by WDs that accrete with intermediate  rates. This is because the effective temperature of low mass rapidly accreting WDs  falls below $\rm 10^{5} K$ as the WD burns at higher rates and its photosphere expands. Comparing the PNLF cut-off value, we find that  WDs with mass of 0.5 $\rm M_{\odot}$ that accrete with intermediate rates can reach magnitudes that extend to -4.3 mag, while the accreting WDs with masses of 0.6 $\rm M_{\odot}$ and accretion rates greater than $0.9 \times 10^{-7}~\rm M_{\odot}~yr^{-1}$  can run into magnitudes as high as -4.8 mag. Another interesting result is that all accreting WDs at all WD masses and accretion rates produce [\ion{O}{III}] magnitudes which are quite similar and many values coincide, suggesting that the accreting WDs models can reproduce the PNLF value for a variety of stellar populations. From these considerations, it follows that young or evolved optically thick PNe powered by a large range of WD masses and accretion rates  within the steady accretion regime can reproduce the PNLF cut-off value.

\section{Discussion and Conclusions}
\label{sec:4}

Motivated by observations and theoretical arguments that mass transfer may take place in binary stars of PNe, in the present work we investigated the optical spectral signatures and evolution of PNe that host a 0.6 $\rm M_{\odot}$ steadily accreting WD at their centres, assuming simple but reasonable assumptions regarding the evolution of the gaseous shell and studying  the case where the nebula remain optically thick for  its whole evolution   \citep[maximum nebula hypothesis][]{2018Gesicki}. For comparison, we included in our simulations   one of the  new evolutionary tracks of \citet{2016bertolammi}  that corresponds to a  WD mass of 0.58 $\rm M_{\odot}$  and metallicity Z=0.02 and we studied how  the observables of PNe that host accreting WDs differ from those powered by single ones. 
Note that different selected metallicity  causes only minor changes to the derived masses and ages of the stars according to the models.

From our simulations, we found that the emission line spectra of PNe powered by accreting WDs do not show significant temporal variations. This is not unexpected, as accreting WDs do not follow a standard post-AGB evolutionary track  and they remain efficient ionizing sources as long as mass accretion occurs. In addition, the non temporal variability of the emission line properties indicates that the final outcome of PNe that host accreting WDs are relatively insensitive to the position and density of the emitting shell.
On the other hand, the nebular line luminosities are heavily dependent on the WDs accretion properties, advocating  that a vast diversity of properties is expected to be met in PNe surrounding steadily-accreting WDs, as each nuclear-burning WD leaves its unique and distinctive imprint on
the surrounding gas.

This is not the case for the single $0.58 \rm M_{\odot}$ model, where the pace of evolution of the nebular luminosities, especially at the early phases, {is consistent  with the variation of its effective temperature and luminosity that follow the evolution of the central star. Consequently after a few thousand of years the PN  shell fades out as its optical line luminosity drops by several orders of magnitude.

These two different evolutionary paths that the two classes of PNe possess result to distinctive features regarding their optical line emission. Whereas during the early PN phase ($t \leq 2000$~yrs) the two optical bright shells reveal similar properties, the persisting ionizing radiation emanating from accreting WDs maintain the nebular line luminosities  almost invariant and inevitable the surrounding shell becomes  up to 4 orders of magnitude brighter than its counterpart powered by the isolated WD.} This result implies that the visibility time and the detection probability of very evolved, extended and dispersed PNe  hosting accreting WDs is much greater. In other words, PNe that expand to a radius of, say 1 pc or more, are more likely to be detected if the central star is an accreting WD than a single one. This result is very important in the context of  general population synthesis studies such as \citet{2006DeMarco}, where the visibility time of PNe in stellar systems is an important parameter required for estimating the size of a PN population and their detectability likelihood.

Additionally, we presented the evolution of some important  emission line ratios of PNe surrounding accreting WDs and we show how they depend on the mass accretion rate. Qualitatively, we found that various  emission line ratios of PNe powered by accreting WDs have a considerable overlap with those powered by a single one during the early evolutionary phases where the single star heats-up towards the turn-around point in the Hertzsprung-Russel diagram as both accreting and single WDs share similar ionizing properties and thus, making the optical spectral signatures of PNe surrounding accreting WDs difficult to be distinctive. However, accreting WDs depending on the mass accretion rate can reveal a much wider range of values that can be a lot lower or higher as compared to those powered by a single counterpart. For evolved PNe, the differences become appreciable and we note the potential usefulness of the emission line ratio [\ion{O}{III}]/$\rm H{\rm \beta}$, which is much greater in PNe hosting accreting WDs and hence, can be used as an index to separate accreting WDs from single ones. 
This combined with the high electron temperature that PNe hosting accreting WDs display at all evolutionary phases, can lend further credence to focusing on these properties in observational searches for accreting WDs' PNe. 
Higher temperature $(\sim 10^{5.5}~ \rm K)$ accreting WDs than those considered in this study may reveal themselves in lower [\ion{O}{III}]/$\rm H{\rm \beta}$ ratio values as the prevailing ionisation stage shifts from $\rm O^{2+}$ to $\rm O^{3+}$. Such objects should also be detectable in X-rays in the absence of significant obscuration.  In any case, the strongest indicator of the presence of an accreting WD in a PN would be the extremely high effective temperature of the central source.
Many PNe  (e.g., M 1-333, MWP~1, NGC 7094, NGC 6058 and NGC 4361 among others) show a series of strong indications that an accreting WD is possibly hosted at their centers \citep{1999vazquez,2009Miranda, 2013Guillen,2015Montez,2019Gonz}. They are physically very extended, their kinematic ages differ from the evolutionary
ages predicted by the models of \citet{2016bertolammi}  and their emission-line ratios and line fluxes are consistent with
ionization by an accreting WD \citep[][]{2009Miranda, 2019Gonz, 2021Gonz}. In addition,
several reveal point like soft X-ray emission from the central
star, the origin of which remains essentially unconstrained while their non-spherical nebular morphologies suggest binarity \citep[][]{2014Freeman, 2015Montez}. However, further observations and detailed photoionization modelling are needed in order to resolve the nature of the sources ionizing such PNe.

In this paper, we also explored the [\ion{O}{III}] brightness of PNe powered by accreting WDs involving in our modeling accreting WD masses in the range of 0.5~-~0.8~$\rm M_{\odot}$ and  various accretion rates within the steady accretion regime. All steadily accreting WDs at all masses and accretion rates are capable to produce very bright [\ion{O}{III}] PNe with absolute magnitude values clustered very close to the PNLF cut-off value. Intriguingly, we found that not only young but also very evolved optically thick PNe can reach that values, suggesting that  the  [\ion{O}{III}] brightness of this class of PNe does not show a strong dependence on their evolutionary state or their central WD mass but are dominantly determined by the mass accretion properties as long as accretion is an ongoing process.
If this scenario is correct, it seems capable to provide a compelling reason why the bright end of PNLF is invariant and does not change with time as it is expected according to the single star evolution models.

Our current analysis -being a first attempt in modeling PNe using an accreting WD as a central source- includes several simplifications, and thus, is far for being considered exhaustive. However, its extracted results advocate towards the binary evolution pathways and mass transfer processes as a key in explaining several observables we receive by a number of PNe including the long standing problem of the invariant PNLF. Thus, tempting motivations are provided for  more thorough studies of accreting WD systems surrounded by PNe, including e.g. a) different PN shell properties,  morphologies and chemical abundances; b) alternative WD accretion models like those of \citet{2013Wolf}, where they use time-dependent calculations and the stable burning boundaries occur at slightly higher effective temperatures, as well as the models of \citet{2014Piersanti}, which include the behavior of He-accreting white dwarfs; c) different accretion regimes involving novae eruptions  and optically-thick accretion winds  and d) direct links with the stellar and binary evolution theories towards the formation of PNe.

\section*{Acknowledgements}
The authors would like to thank the referee, Prof. Krzysztof M. Gęsicki for
 thorough comments that improved the manuscript.
This research is co-financed by Greece and the European Union (European Social Fund-ESF) through the Operational Programme “Human Resources Development, Education and Lifelong Learning 2014-2020” in the context of the project “On the interaction of Type Ia Supernovae with Planetary Nebulae” (MIS 5049922). A.C. acknowledge the support of this work by the project ``PROTEAS II'' (MIS 5002515), which is implemented under the Action ``Reinforcement of the Research and Innovation Infrastructure'', funded by the Operational Programme ``Competitiveness, Entrepreneur- ship and Innovation'' (NSRF 2014–2020) and co-financed by Greece and the European Union (European Regional
Development Fund). D.J. acknowledges support from the Erasmus+ programme of the European Union under
grant number 2020-1-CZ01-KA203-078200. PJG and OM are supported by NRF SARChI grant 111692.

%%%%%%%%%%%%%%%%%%%%%%%%%%%%%%%%%%%%%%%%%%%%%%%%%%
\section*{Data Availability}

The data underlying this article will be shared on reasonable request to the corresponding author.



\bibliographystyle{mnras}
\bibliography{accwdspne} % if your bibtex file is called example.bib

%\begin{thebibliography}{99}
%\end{thebibliography}

%%%%%%%%%%%%%%%%%%%%%%%%%%%%%%%%%%%%%%%%%%%%%%%%%%

%%%%%%%%%%%%%%%%% APPENDICES %%%%%%%%%%%%%%%%%%%%%

%\appendix



%%%%%%%%%%%%%%%%%%%%%%%%%%%%%%%%%%%%%%%%%%%%%%%%%%

% Don't change these lines
\bsp	% typesetting comment
\label{lastpage}

\end{document}